\documentclass[twocolumn,showpacs,preprintnumbers,amsmath,amssymb]{revtex4}

\usepackage{epsfig}
\usepackage{epstopdf}
\usepackage{graphicx}
\usepackage{dcolumn}
\usepackage{bm}
\usepackage[usenames]{color}

\begin{document}

\title{Rigid motion revisited: rigid quasilocal frames}

\author{Richard~J.~Epp}
\email{rjepp@sciborg.uwaterloo.ca}
\author{Robert~B.~Mann}
\email{rbmann@sciborg.uwaterloo.ca}
\author{Paul~L.~McGrath}
\email{pmcgrath@phys.ualberta.ca}
\affiliation{Department of Physics and Astronomy\\University of
Waterloo\\Waterloo, Ontario, Canada}

\date{\today}

\begin{abstract}
In this first of a series of papers we will introduce the notion of a {\it rigid quasilocal frame} (RQF) as a geometrically natural way to define a ``system" in the context of the dynamical spacetime of general relativity.  An RQF is defined as a two-parameter family of timelike worldlines comprising the worldtube boundary (topologically $\mathbb{R}\times \text{S}^2$) of the history of a finite spatial volume, with the rigidity conditions that the congruence of worldlines is expansion-free (the ``size" of the system is not changing) and shear-free (the ``shape" of the system is not changing).  This definition of a system is anticipated to yield simple, exact geometrical insights into the problem of motion in general relativity.  It begins by answering, in a precise way, the questions {\it what} is in motion (a rigid two-dimensional system boundary with topology $\text{S}^2$, and whatever matter and/or radiation it happens to contain at the moment), and what motions of this rigid boundary are possible.  Nearly a century ago Herglotz and Noether showed that a three-parameter family of timelike worldlines in Minkowski space satisfying Born's 1909 rigidity conditions does not have the six degrees of freedom we are familiar with from Newtonian mechanics, but a smaller number---essentially only three. This result curtailed, to a large extent, subsequent study of rigid motion in special and (later) general relativity.  We will argue that in fact we {\it can} implement Born's notion of rigid motion in both flat spacetime (this paper) and arbitrary curved spacetimes containing sources (subsequent papers)---with precisely the expected three translational and three rotational degrees of freedom (with arbitrary time dependence)---provided the system is defined {\it quasilocally} as the two-dimensional set of points comprising the {\it boundary} of a finite spatial volume, rather than the three-dimensional set of points within the volume.
\end{abstract}


\maketitle

\section{Introduction}

Rigid motion in Newtonian space-time has six degrees of freedom: three translations and three rotations.  In other words, there are six arbitrary time-dependent degrees of freedom in constructing a three-parameter congruence of ``timelike" worldlines such that the distance between each pair of infinitesimally separated worldlines remains constant.  This fact greatly simplifies the description of the motion of (rigid) extended bodies, and motivated M.~Born \cite{Born1909} to propose a similar definition of rigid motion in the context of special relativity, with ``distance" now defined as the orthogonal distance between neighboring worldlines measured with the Minkowski line element.  Soon afterwards, G.~Herglotz~\cite{Herglotz1910} and F.~Noether~\cite{Noether1910} proved that such Born-rigid motions exist, but they have essentially only three degrees of freedom.  More precisely, there are two types of Born-rigid motion in special relativity: (1) arbitrary time-dependent translations with {\it no} rotation (so-called {\it plane} motions), and (2) motions generated by a Killing field, i.e., the repeated action of one element of the Poincar\'{e} group (so-called {\it group} motions).\cite{Salzman+Taub1954,Eriksen+Mehlen1982}  As the simplest representative example of the latter, the only possible motion of a Born-rigid body with one point fixed is an eternal unchangeable rotation.\cite{Eriksen+Mehlen1982}  Born-rigid motion in the context of general relativity, especially the rotating case, is considerably more subtle---see \cite{Mason+Pooe1987} and references therein.

A body of literature has grown out of exploring various relaxations or modifications of Born's notion of rigidity (for examples, see \cite{Llosa+Soler2004,Llosa+Soler2000,Bel+Llosa1995,Bel+Martin+Molina1994,Bona1982}), but as far as the authors are aware no proposal has emerged that recovers the full set of six arbitrary time-dependent degrees of freedom in a geometrically natural (coordinate independent) way.

In this paper we will introduce the notion of a {\it rigid quasilocal frame} (RQF), which is simply Born's notion of rigidity applied not to a {\it three}-parameter congruence (history of a spatial volume-filling set of points), but to a {\it two}-parameter congruence (history of the set of points on the surface bounding a spatial volume).  This volume-to-surface, or quasilocal, relaxation provides a simple and geometrically natural way around the restrictions found by Herglotz and Noether: Whereas Born-rigidity for a three-parameter congruence involves six differential constraints on three functions (an over determined system),\cite{Llosa+Soler2000} we will see that an RQF involves only {\it three} differential constraints on three functions, and argue that the space of solutions is parameterized by precisely six arbitrary time-dependent degrees of freedom.  Remarkably, the existence of these degrees of freedom is intimately connected with the well known fact that a two-sphere (as opposed to a closed two-surface of any other genus), regardless of its geometry (the size and shape of the rigid ``box" bounding the volume), always admits precisely six conformal Killing vectors, which generate an action of the Lorentz group on the sphere: three rotations and three boosts.\cite{Held+Newman+Posadas1970}  A single observer undergoing arbitrary acceleration and rotation can be thought of as being acted upon by a time-dependent sequence of local Lorentz transformations.  In essence, an RQF extends this notion to a two-sphere's worth of observers being acted upon by a time-dependent sequence of ``quasilocal Lorentz transformations."

The paper is organized as follows.  In {\S}\ref{Definition} we define the notion of an RQF in a general (3+1)-dimensional spacetime.  In {\S}\ref{SimpleExamples} we construct two simple, representative examples of RQFs in flat spacetime: (1) a round sphere undergoing arbitrary time-dependent translations, with no rotation, and (2) a round sphere undergoing {\it constant} rotation, with no translation. These examples illustrate the two types of rigid motion allowed by the Herglotz-Noether theorem.  In {\S}\ref{ArbitraryPerturbations} we begin to go beyond these types by considering arbitrary time-dependent infinitesimal perturbations about the simplest RQF---a non-accelerating and non-rotating round sphere in flat spacetime.  We demonstrate that the tangent space to the RQF solution space, at the point of this simplest solution, is spanned by precisely six arbitrary functions of time and, moreover, establish the connection between these degrees of freedom and the natural action of the Lorentz group on the sphere, mentioned above.  We close {\S}\ref{ArbitraryPerturbations} with a consideration of infinitesimal perturbations about a generic RQF in curved spacetime, which reveals a peculiar ``nonlocality" in time inherent in RQFs with finite time-dependent rotation.

Indeed, this is where the real difficulty lies: constructing, even in flat spacetime, RQFs with arbitrary finite time-dependent rotation.  The reason can be traced to the relativity of simultaneity, which has the most severe consequences for congruences with {\it twist}, i.e., rotating systems, for which the congruence is not hypersurface orthogonal.  As we shall see, this necessarily activates a certain nonlinear term in the rigidity equations that involves a time derivative, changing the basic nature of the partial differential equations involved.  Thus the simplest example that would nevertheless provide a strong proof of principle is a flat spacetime RQF undergoing an arbitrary finite time-dependent rotation, with no translation. (This problem is essentially the quasilocal analogue of the well known ``Ehrenfest's paradox," in which a rigid body at rest can never be brought into uniform rotation.\cite{EinsteinStudies-RotatingDisk})  In {\S}\ref{GeneralRotation} we construct precisely such an example, solving the rigidity equations iteratively in powers of the rotation rate and its time derivatives.  Computing the first few terms in the series we find that our approximate solution can be pushed with confidence to the rather extreme case of a round sphere RQF spinning up from rest to angular velocities for which observers on the sphere's equator are moving at 1/3 the speed of light, on a time scale less than the time it takes the sphere to rotate a small fraction of one revolution.  Finally, in {\S}\ref{Conclusions} we present conclusions.

\section{Definition of an RQF}\label{Definition}

We will begin by introducing some notation.  Let $M$ be a smooth four-dimensional manifold endowed with a Lorentzian spacetime metric, $g_{ab}$, with signature $+2$.  Naturally associated with $g_{ab}$ is its torsion-free, metric-compatible covariant derivative operator, $\nabla_{a}$, and volume element $\epsilon_{abcd}$.  Let $\mathcal B$ denote a two-parameter family of timelike worldlines with topology $\mathbb{R}\times \text{S}^2$, i.e., a timelike worldtube that represents the history of a two-sphere's-worth of observers bounding a finite spatial volume.  Let $u^a$ be the future-directed unit vector field tangent to this congruence, representing the observers' four-velocity.  The spacetime metric, $g_{ab}$, induces on $\mathcal B$ a spacelike outward-directed unit normal vector field, $n^{a}$, and a Lorentzian three-metric, $\gamma_{ab}:=g_{ab}-n_{a}n_{b}$.  At each point $p\in {\mathcal B}$ we have a {\it horizontal} subspace, $H_p$, of the tangent space to $\mathcal B$ at $p$, consisting of vectors orthogonal to both $u^a$ and $n^a$.  Let $\sigma_{ab}:=\gamma_{ab}+u_a u_b$ denote the spatial two-metric induced on $H:=\bigcup_p H_p$. Finally, let $\epsilon_{ab} := \epsilon_{abcd}u^c n^d$ denote the corresponding volume element associated with $H$.  The time development of our congruence is described by the tensor field $\theta_{ab}:=\sigma_a^{\phantom{a}c}\sigma_b^{\phantom{b}d}\nabla_c u_d$.  We adopt the usual terminology: the {\it expansion} is $\theta := \sigma^{ab}\theta_{ab}$ (the
trace part); the {\it shear} is $\theta_{< ab >} :=
\theta_{(ab)}-{1\over2}\theta\sigma_{ab}$ (the symmetric trace-free
part, here and elsewhere denoted by angle-brackets); and the {\it twist} is $\nu:=\frac{1}{2}\epsilon^{ab}\theta_{ab}$ (the
antisymmetric part).

A {\it rigid quasilocal frame} is defined as a congruence of the type just described, with the additional conditions that the expansion and shear both vanish, i.e., the size and shape, respectively, of the boundary of the finite spatial volume---as seen by our observers, do not change with time:
\begin{equation} \label{eq:RigidityCondition}
\theta = 0 =
\theta_{<ab>}\;\;\;\Longleftrightarrow\;\;\; \theta_{(ab)}=0 .
\end{equation}
These three differential constraints ensure that $\sigma_{ab}$ is a well defined two-metric on the quotient space of the congruence, $\mathcal Q\simeq \text{S}^2$, i.e., the space of the observers' worldlines.  It describes the intrinsic geometry of the rigid ``box" bounding the volume, as measured locally by our two-sphere's-worth of observers.  Notice that there is no restriction on $\nu$---the twist of the congruence---since we want to allow for the possibility of our rigid box to rotate, in which case the subspaces comprising $H$ are not integrable, i.e., $u^a$ is not hypersurface orthogonal as a vector field in $\mathcal B$.

It is instructive at this point to recall the Landau-Lifshitz radar ranging spatial metric associated with a three-parameter family of observers moving along a congruence formed by the integral curves of a time flow vector field, $\xi^a$.[14]  With the observers' (unit) four-velocity given by $u^a := \xi^a/N$, the Landau-Lifshitz metric is defined as $h_{ab}:= g_{ab}+u_a u_b$, which clearly measures the---in general time-dependent---orthogonal distance between infinitesimally separated observers.  Clearly $\sigma_{ab}:=\gamma_{ab}+u_a u_b$ (with $u^a := \xi^a/N$ tangent to $\mathcal B$) is the two-dimensional analogue of $h_{ab}$ for a timelike congruence in the three-dimensional spacetime $({\mathcal B},\gamma_{ab})$.  Landau and Lifshitz state that for $h_{ab}$ to be well defined on the quotient space of the congruence it is necessary (and obviously sufficient) that $\xi^a$ be a (timelike) Killing vector field in the four-dimensional spacetime $(M,g_{ab})$.  The corresponding condition in the three-dimensional spacetime $({\mathcal B},\gamma_{ab})$ reads: $\gamma_{(a}^{\phantom{(a}c} \gamma_{b)}^{\phantom{b)}d}\nabla_c \xi_d=0$.  However, we observe that in both the four- and three-dimensional cases this condition is overly restrictive, and so unnecessarily rules out most interesting spacetimes.  Only the {\it spatial projection} of this condition is necessary and sufficient for the respective radar ranging metrics to be well defined on the respective quotient spaces: $h_{(a}^{\phantom{(a}c} h_{b)}^{\phantom{b)}d}\nabla_c \xi_d=0$ in the four-dimensional case and $\sigma_{(a}^{\phantom{(a}c} \sigma_{b)}^{\phantom{b)}d}\nabla_c \xi_d=0$ in the three-dimensional case.  Because of the spatial projection we can replace $\xi^a$ with $u^a$ in both cases, and what emerges is the Born rigidity condition ($\theta_{(ab)}=0$ in the three-dimensional case).  We thus see precisely how the Born rigidity condition is weaker than the Killing vector condition.  There are two points to make in this regard.  First, as noted in the Introduction, the Born rigidity condition is still too restrictive to be of much interest in the four-dimensional case, but not necessarily in the three-dimensional (RQF) case.  Thus, to the extent that RQFs exist, they might be thought of as the closest we can get to a timelike Killing vector field in a general, dynamical spacetime.  Restricting our attention now to the three-dimensional case, our second point is that under the Killing vector condition the observers' proper acceleration ($a^a :=u^b \nabla_b u^a$) projected into $\mathcal B$ ($\alpha^a := \gamma^a_{\phantom{a}b}a^b = \sigma^a_{\phantom{a}b}a^b$) is just the gradient of a scalar: $\alpha_a = \sigma_a^{\phantom{a}b}\nabla_b \ln N$.  However, under the Born rigidity condition more general accelerations are allowed.  As we shall see shortly (in the context of equation~(\ref{eq:ObserversAcceleration}) below), the added generality represents---with a small caveat---precisely the dynamical degrees of freedom of the intrinsic geometry of RQFs.

To further clarify the RQF construction, and to establish notation for the examples in subsequent sections, let us restore the speed of light, $c$, (which was hitherto set to 1) and introduce a coordinate system adapted to the congruence.  Thus, let two functions $x^i$ on ${\mathcal B}$ locally label the observers, i.e., the worldlines of the congruence. Let $t$ denote a ``time" function on ${\mathcal B}$ such that the surfaces of constant $t$ form a foliation of ${\mathcal B}$ by two-surfaces with topology $\text{S}^2$.  Collect these three functions together as a coordinate system, $x^\mu := (t,x^i)$, and set $u^\mu := N^{-1}\delta_t^\mu$, where $N$ is a lapse function ensuring that $u\cdot u=-c^2$. The general form of the induced metric $\gamma_{ab}$ then has adapted coordinate components:
\begin{equation} \label{eq:InducedMetric}
\gamma_{\mu\nu} = \left(
\begin{array}{cc}
-c^2N^2 & Nu_j \\
Nu_i & \sigma_{ij}-\frac{1}{c^2}u_i u_j
\end{array}
\right).
\end{equation}
Here $\sigma_{ij}$, and the shift covector $u_i$, are the $x^i$ coordinate components of $\sigma_{ab}$ and $u_{a}$, respectively.  (We remind the reader that because $u^\mu := N^{-1}\delta_t^\mu$ instead of $u_\mu := -N\delta^t_\mu$, this is {\em not} an ADM decomposition of $\gamma_{ab}$.)  Note that $\sigma_{ij}\,dx^i\,dx^j$ is the radar ranging, or orthogonal distance between infinitesimally separated pairs of observers' worldlines, and it is a simple exercise to show that the RQF rigidity conditions in equation~(\ref{eq:RigidityCondition}) are equivalent to the three conditions $\partial\sigma_{ij}/\partial t =0$.  The resulting time-independent $\sigma_{ij}$ is the metric induced on ${\mathcal Q}\simeq \text{S}^2$.

In other words, an RQF is a {\it rigid} frame in the sense that each observer sees himself to be permanently at rest with respect to his nearest neighbours.  The idea is that this is true even if, for example, a gravitational wave is passing through the RQF, in which case neighboring observers must undergo different proper four-accelerations in order to maintain nearest-neighbour rigidity.  They will also, in general, observe different precession rates of inertial gyroscopes.  Indeed, these inertial accelerations and rotations encode information about both the motion of their rigid box and the nontrivial nature of the spacetime it is immersed in.

It is not obvious that the rigidity conditions (\ref{eq:RigidityCondition}) can, in general, be satisfied.  In this paper we will explore this possibility in the simplest possible context of RQFs in flat spacetime, and in subsequent papers will consider more general spacetimes, including those representing gravitational waves.  However, assuming that these conditions {\it are} satisfied, we are then free to perform a time-{\it in}dependent coordinate transformation amongst the $x^i$ (a relabeling of the observers) such that $\sigma_{ij}$ takes the form $\sigma_{ij}=\Omega^2 \, \mathbb{S}_{ij}$, where $\Omega^2$ is a time-independent conformal factor encoding the size and shape of the rigid box, and $\mathbb{S}_{ij}$ is the standard metric on the unit round sphere. For example, if the observers' two-geometry is a round sphere of area $4\pi r^2$, and the observers are labeled by the standard spherical coordinates $x^i=(\theta,\phi)$, then $\mathbb{S}_{ij}=\text{diagonal}(1,\sin^2\theta)$ and $\Omega=r$.  We are also free to change the time foliation of ${\mathcal B}$ such that $N=1$, i.e., $t$ is proper time for the observers.

Thus we see that the intrinsic three-geometry of an RQF has two functional degrees of freedom that---with the choice of coordinate-fixing described above---are encoded in the two components of the shift covector field, $u_j$ (which are functions of $t$ and $x^i$), as well as the time-{\it in}dependent conformal factor, $\Omega$, encoding our choice of size and shape of the rigid box.  We may also think of the dynamical degrees of freedom, $u_j$, as being encoded in the observers' (coordinate independent) proper acceleration tangential to $\mathcal B$ ($\alpha^{a}$ defined earlier), whose covariant components are
\begin{equation}\label{eq:ObserversAcceleration}
\alpha_{j}={1\over N}\,\dot{u}_{j}+c^2\partial_{j}\ln N,
\end{equation}
in the adapted coordinate system.  (Here an over-dot denotes partial derivative with respect to $t$, and $\partial_{j}:=\partial/\partial x^{j}$.)  More precisely, in addition to $\alpha_j$ we are free to specify the twist, $\nu$, on one cross section of $\mathcal B$, where
\begin{equation}\label{eq:nu}
\nu = \frac{1}{2}\epsilon^{ij}(\partial_i u_j - \frac{1}{c^2}\alpha_i u_j ),
\end{equation}
and $\epsilon^{ij}$ are the $x^i$ coordinate components of $\epsilon^{ab}$.

A full discussion of the intrinsic and extrinsic geometry of RQFs---their kinematics and dynamics, respectively, will be given elsewhere.  Our goal here is only to construct some representative examples of RQFs, and argue that RQFs have the same degrees of freedom of motion as a Newtonian rigid
body.  (Note that $u_j$, the two dynamical functional degrees of freedom in the RQF intrinsic three-geometry, should not be confused with the six time-dependent functions describing the rigid motion, although they are related.  As will become clearer in the examples below, the latter determine the congruence, and hence both the intrinsic and extrinsic three-geometry of $\mathcal B$ as a hypersurface embedded in an ambient four-geometry.  As intimated earlier in this section, the intrinsic and extrinsic three-geometry in turn contain information about gravitational and other fluxes of energy, momentum and angular momentum through the system boundary.  These fluxes will not be discussed in this paper.)

\section{\label{SimpleExamples}Simple examples of RQFs}

We will construct two representative examples of RQFs in flat spacetime: (1) a round sphere undergoing arbitrary time-dependent translations, with {\it no} rotation, and (2) a round sphere rotating at a {\it constant} rate, with no translation.

\subsection{\label{SimpleExamples-TranslationOnly}Translation Only}

For this example we let $X^a =(cT,X,Y,Z)$ denote Minkowski coordinates in an inertial reference frame in flat spacetime, with metric $g_{ab}=\text{diagonal}(-1,1,1,1)$.  Let $X^a = \xi^a (t)$ define an arbitrary timelike worldline ${\mathcal C}_0$, parameterized by proper time, $t$, around which we will construct the timelike worldtube, $\mathcal B$, of our accelerating RQF.  Let $U^a =  \dot{\xi}^a$ be the four-velocity along ${\mathcal C}_0$, such that $U\!\cdot\! U = -c^2$, and define the timelike unit vector $e_{0}^{\phantom{0}a}:=U^a /c$.  At some point along ${\mathcal C }_0$, say $t=0$, choose a spatial triad $e_{I}^{\phantom{I}a}$, $I=1,2,3$, orthogonal to $U^a$.  Define $e_{I}^{\phantom{I}a}$ all along ${\mathcal C}_0$ by Fermi-Walker transport (no rotation of the spatial triad):\cite{MTW1973}
\begin{equation}\label{FermiWalker}
\nabla_{e_{0}}e_{A}^{\phantom{A}a}=-\Omega^{a}_{\phantom{a}b}\,e_{A}^{\phantom{A}b}.
\end{equation}
Here we have collected $e_{0}^{\phantom{0}a}$ and $e_{I}^{\phantom{I}a}$ into a tetrad, $e_{A}^{\phantom{A}a}$, defined along ${\mathcal C}_0$, and defined $\Omega^{ab}:=\frac{1}{c^3}(A^a U^b - U^a A^b )$, where $A^a:=U^b \nabla_b U^a=\dot{U}^a$ is the acceleration along ${\mathcal C}_0$ (and of course $U\!\cdot\! A=0$).  In particular, from equation~(\ref{FermiWalker}) we have $\dot{e}_{I}^{\phantom{I}a}=\frac{1}{c^2}A_{I}U^a$, where $A_{I}:=e_{I}^{\phantom{I}a}A_a$ are the triad components of the proper acceleration of ${\mathcal C}_0$.

Let us now embed, in our spacetime, a two-parameter family of timelike worldlines around ${\mathcal C}_0$:
\begin{equation}\label{eq:AccelerationOnly}
X^a = \Xi^a(t,\theta,\phi):=
\xi^a(t)+r\,r^I(\theta,\phi)\,e_I^{\;\;a}(t),
\end{equation}
representing a two-sphere's worth of observers (labeled with spherical coordinates $\theta$, $\phi$) comprising the timelike worldtube, $\mathcal B$.  Here $r^I(\theta,\phi):=(\sin\theta\cos\phi, \sin\theta\sin\phi, \cos\theta)$ are the standard direction cosines of a radial unit vector in spherical coordinates in Euclidean 3-space, and $r$ is a variable parameter that will turn out to be the areal radius of our round sphere RQF.  A simple calculation reveals that, in the adapted coordinate system $x^\mu=(t,\theta,\phi)$, the components of the metric induced on $\mathcal B$ have the form of equation (\ref{eq:InducedMetric}) with $N=1+\frac{r}{c^2}\,r^I A_I$, $u_i=0$, and $\sigma_{ij} = r^2\,\mathbb{S}_{ij}$, where $\mathbb{S}_{ij}$ is the unit round sphere metric introduced in the previous section.  The components of the observers' proper acceleration tangential and normal to $\mathcal B$ are then easily found to be
\begin{eqnarray}
\alpha_j &=& {1\over N}\,r \,  A_{I}\,\partial_j r^{I},\\
n\cdot a &=& {1\over N}\,r^{I}A_{I},
\end{eqnarray}
and obviously the twist, $\nu$, vanishes.  Note that the normal acceleration, $n\cdot a$, is a component of the extrinsic curvature of $\mathcal B$.  Insofar as we are not developing the formalism to analyze extrinsic curvature in this paper, the result is stated for completeness, but without proof (although the result is clearly plausible in this simple case).

Thus we have constructed an RQF that depends on three arbitrary functions of time: the three independent components of $\xi^{a}(t)$, or, if you will, the three components of the acceleration, $A_{I}(t)$, which describes a rigid round sphere of area $4\pi r^2$ undergoing arbitrary time-dependent translations.  Despite the proper accelerations the observers experience, both tangential and normal to the spherical frame they define, they may consider themselves to be ``stationary" in the sense described earlier: each observer sees himself to be permanently at rest with respect to his nearest neighbours.  In the spirit of Einstein's principle of equivalence\cite{EinsteinStudies-EquivalencePrinciple} they can consider themselves to be at rest in a time-dependent gravitational field that varies in strength and direction from one observer to the next.

There are two points worth noting.  First, observe that our construction is valid only if $N  >  0$, i.e., $\text{max}\{|A_I(t)|\}<\frac{c^2}{r}$.  In other words, our RQF is restricted in size by a ``Rindler horizon." This is not surprising, and obviously must be a generic property of RQFs: for a given size of bounding box there must be a maximum value of some acceleration parameter (in this case $A_I$) in order that the proper acceleration of each observer remain finite.  To the extent that RQFs provide a general description of physical systems (to be discussed in subsequent papers), we speculate that quantization may introduce a minimum size for RQFs, and hence a maximum acceleration in quantum gravity.  Second, observe that there is what might aptly be called a ``temporal stress" associated with the fact that different observers require different proper accelerations to ensure rigidity, and thus different observers' clocks record proper time at different relative rates.  This is analogous to the well known fact that the back of a rocket must have a greater proper acceleration than the front for the rocket to maintain constant proper length (rigidity) as measured by co-moving observers, and that these observers then necessarily experience a ``temporal stress."\cite{Giannoni+Gron1978}

\subsection{\label{SimpleExamples-ConstantRotationOnly}Constant Rotation Only}

For this example we let $X^a =(cT,P,\Phi,Z)$ denote cylindrical coordinates in an inertial reference frame in flat spacetime, with metric $g_{ab}=\text{diagonal}(-1,1,P^2,1)$.  As in the previous example, we wish to embed a two-parameter family of timelike worldlines, labeled by coordinates $x^i=(\theta,\phi)$ and representing observers on a rigid round sphere of areal radius $r$, but this time with each observer rotating with constant angular velocity $\omega$ about the $Z$-axis ($\omega$ as measured by observers at rest in the inertial reference frame). Beginning with the ansatz:
\begin{equation}\label{eq:ConstantRotationAnsatz}
\begin{array}{lll}
T & = & t\\
P & = & \rho(\theta)\\
\Phi & = & \phi + \omega t\\
Z & = & z(\theta),
\end{array}
\end{equation}
a simple calculation yields an induced metric of the form given in equation (\ref{eq:InducedMetric}), with
\begin{eqnarray}
N & = & \sqrt{1-\omega^2\rho^2/c^2}\\
u_i & = & (\,0\,,\,\omega\rho^2/N\,)\\
\sigma_{ij} & = &
\left(
\begin{array}{cc}
\rho^{\prime \, 2}+z^{\prime \, 2} & 0 \\
0 & \rho^2/N^2
\end{array}
\right),
\end{eqnarray}
where a prime denotes differentiation with respect to $\theta$.  Evidently, for the rotating observers to see a round sphere of areal radius $r$ we require:
\begin{eqnarray}
\rho(\theta) & = & {{r\sin\theta}\over{\sqrt{1+\gamma^2\sin^2\theta}}}\label{eq:rho}\\
z(\theta) & = & -r \int_{\frac{\pi}{2}}^{\theta}\mathrm{d}\tilde{\theta}\, \sqrt{1-\frac{\cos^2\tilde{\theta}}{(1+\gamma^2\sin^2\tilde{\theta})^3}},\label{eq:z}
\end{eqnarray}
where $\gamma:=r\omega/c$ is a dimensionless measure of how relativistic the system is.  Unfortunately, the integral for $z(\theta)$ cannot be expressed in terms of elementary functions.  Figure~1 shows the results of numerical integration for $r=1$ and three values of $\gamma$.  The sphere, which is round for our co-rotating observers, appears to inertial observers as an increasingly cigar-shaped surface as $|\gamma |$ increases.

To understand this figure, consider the tangential velocity of observers on the equator ($\theta=\pi/2$), given by
\begin{equation}\label{eq:vEquator}
\frac{v_\text{equator}}{c}=\frac{1}{c}\,\omega\,\rho(\tfrac{\pi}{2})=
\frac{\gamma}{\sqrt{1+\gamma^2}}.
\end{equation}
(Observe that $\gamma$, and hence the angular velocity, $\omega$, can range from $-\infty$ to $+\infty$.)  Recalling Einstein's famous rotating disk thought experiment,\cite{EinsteinStudies-RotatingDisk} in which rotating observers on the edge of the disk measure a greater circumference than inertial observers, the radius of the equator of our sphere must contract as $|\omega |$ increases in order to maintain the desired circumference of $2\pi r$.  Also note that, since there is no length contraction in a $\phi=$ constant plane, the length of each of the curves in figure~1 is simply $\pi r$.  Thus, in the ultra relativistic limit $|\omega|\rightarrow\infty$, $z(\theta)$ ranges from $-\pi r/2$ to  $+\pi r/2$.

\begin{figure}
\centering
\includegraphics[width=0.5\textwidth]{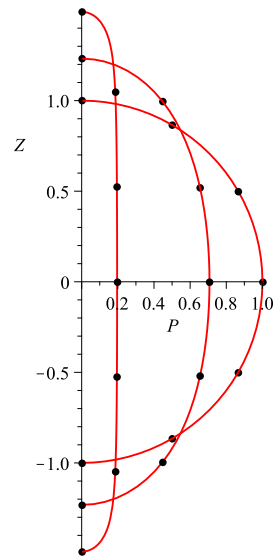}
\caption{The shape of the rotating observers' round sphere as seen by observers in the inertial reference frame: plots of the parametric curve $P=\rho(\theta)$, $Z=z(\theta)$ for $r=1$ and $\gamma =$ 0, 1 and 5.  The last case corresponds to $v_\text{equator}\approx 0.98\,c$. The dots show the locations of observers with $\theta=0,\pi/6,2\pi/6,\ldots,\pi$.  Since there is no length contraction in a $\phi=$ constant plane, the dots are equally spaced along each curve.}
\end{figure}

Substituting these results into
equations~(\ref{eq:ObserversAcceleration}) and (\ref{eq:nu}) we find
\begin{eqnarray}
\alpha_\theta &=&
-\frac{r^2\omega^2\sin\theta\cos\theta}{1+\gamma^2\sin^2\theta}, \quad \alpha_\phi=0\\
\nu &=& \frac{\omega\cos\theta}{\sqrt{1+\gamma^2\sin^2\theta}}.
\end{eqnarray}
Notice that the magnitude of the observers' tangential proper acceleration along the lines of longitude is maximum around mid latitudes and is directed towards the poles, and that the magnitude of the twist of the congruence is maximum at the poles, as one might expect.  As in the previous example, for completeness we state, without proof, that the component of the observers' proper acceleration normal to $\mathcal B$ turns out to be
\begin{equation}
n\cdot a=-{\frac {r {\omega}^{
2}  \sin^2 \theta   \sqrt {
 1+3\gamma^2+3\gamma^4
\sin ^{2}  \theta +\gamma^{6} \sin^4 \theta  }}{1+\gamma^2 \sin^2  \theta }},
\end{equation}
which in the slow rotation limit reduces to $-r\omega^2\sin^2\!\theta$ as expected.  Observe that even in this very simple example of constant rotation, compared to the arbitrary time-dependent acceleration in the previous example, the expressions for acceleration (and twist) are significantly more complicated here.  In the context of relativistic rigidity, rotation is inherently much more subtle than translation.

\section{\label{ArbitraryPerturbations}RQFs have six degrees of freedom}

In the previous section we wrote down the general solution for a round sphere RQF in flat spacetime undergoing arbitrary acceleration but no rotation.  In addition, we saw that even {\it constant} rotation is considerably more complicated, and anticipate that any closed form solution for time-{\it dependent} rotation, which is of the most interest to us, may well be intractable.  Thus, our strategy in this section is to consider an arbitrary {\it infinitesimal} perturbation about the trivial solution---a
non-accelerating, non-rotating round sphere RQF in flat spacetime, which is described by equation~(\ref{eq:AccelerationOnly}) with $A_{I}=0$.  Perturbing about this solution we begin with the ansatz:
\begin{equation}\label{eq:GeneralPerturbation}
X^a =
\xi^a+r ( r^I +\lambda f^{I}) e_I^{\phantom{I}a}+  \lambda r f^{0}e_0^{\phantom{0}a},
\end{equation}
where $\lambda$ is an infinitesimal parameter and $f^{I}(t,\theta,\phi)$ are three arbitrary functions allowing complete freedom for the observers to ``wiggle around."  A fourth arbitrary function, $f^{0}(t,\theta,\phi)$, is added to allow for a change in the time foliation of $\mathcal B$.  While the RQF rigidity conditions are obviously invariant under such a time reparametrization (which manifests itself in $\sigma_{ij}$ being independent of  $f^0$ in equation~(\ref{eq:PerturbedSigma}) below), a particular choice of $f^{0}$ will prove convenient later.  In our calculations we will retain only terms linear in $\lambda$.  The goal is to determine the dimension of the RQF solution space at the trivial solution point.  Under the fairly mild assumption that the dimension is a continuous function on the solution space, we will thus determine the dimension of the RQF solution space in general, both for flat and curved spacetimes.

Calculating the induced metric as before we now find:
\begin{eqnarray}
N & = & 1+\lambda \frac{r}{c}\dot{f}^0  +O(\lambda^2)\label{eq:PerturbedN}\\
u_i & = & \lambda r^2 ( \dot{f}_{i} - \frac{c}{r}\partial_{i}f^{0}) + O(\lambda^2)\label{eq:Perturbedu_i}\\
\sigma_{ij} & = & r^2\,[\,(1+2\lambda F)\,\mathbb{S}_{ij} + 2\lambda \mathbb{D}_{(i}f_{j)}]+ O(\lambda^2).\label{eq:PerturbedSigma}
\end{eqnarray}
Here we have made the decomposition
\begin{equation}\label{eq:fIdecomposition}
f^{I}(t,\theta,\phi)=F(t,\theta,\phi)\,r^{I}(\theta,\phi) + f^{i}(t,\theta,\phi)\,\mathbb{B}_{i}^{I}(\theta,\phi),
\end{equation}
where $\mathbb{B}_{i}^{I}(\theta,\phi) := \partial_i r^{I}(\theta,\phi)$ and  $f_{i}:=\mathbb{S}_{ij}f^{j}$, and we have made use of the completeness relation $\delta^{IJ}=r^{I}r^{J}+\mathbb{S}^{ij} \mathbb{B}_{i}^{I}\mathbb{B}_{j}^{J}$, where $\mathbb{S}^{ij}$ is the matrix inverse of $\mathbb{S}_{ij}$.  Finally, $\mathbb{D}_i$ appearing in equation~(\ref{eq:PerturbedSigma}) is the covariant derivative operator associated with the unit round sphere metric, $\mathbb{S}_{ij}$.

We now demand that $\sigma_{ij}=r^2\,\mathbb{S}_{ij}$, i.e., the same rigid box we began with (a round sphere of areal radius $r$), but possibly in a state of motion different from rest.  Taking the trace and trace-free parts of equation~(\ref{eq:PerturbedSigma}) yields three linear partial differential equations
\begin{eqnarray}
F & = & -\frac{1}{2}\,\mathbb{D}\!\cdot\! f\label{eq:Fequation}\\
\mathbb{D}_{<i}f_{j>}&=&0\label{eq:CKVequation}
\end{eqnarray}
in the three unknown functions $F$ and $f^i$.  The first equation tells us that $F$ (the radial perturbation) is determined by the vector field $f^i$ (the tangential perturbation), and the second tells us that $f^i$ must be a conformal Killing vector (CKV) field on the unit round sphere.  It is well known that any two-surface with the topology $S^2$ admits precisely six CKVs (compared with two CKVs for a torus, and zero CKVs for any closed surface of higher genus\cite{Nakahara}), and as generators of infinitesimal diffeomorphisms they form a representation of the Lorentz algebra.\cite{Held+Newman+Posadas1970}

To construct them explicitly, let $\mathbb{E}_{ij}$ denote the volume element associated with $\mathbb{S}_{ij}$.  Taking the three functions $r^I$ as a basis for the $\ell=1$ spherical harmonics we construct two sets of $\ell=1$ spherical harmonic covector fields: three {\it boost} generators, $\mathbb{B}^I_i := \mathbb{D}_i r^I$ (which were defined above), and three {\it rotation} generators, $\mathbb{R}^I_i := -\mathbb{E}_{i}^{\phantom{i}j}\mathbb{B}_{j}^I$.  The contravariant form of these generators is given by $\mathbb{B}_I^i:=\delta_{IJ}\mathbb{S}^{ij}\mathbb{B}^J_j$ and $\mathbb{R}_I^i:=\delta_{IJ}\mathbb{S}^{ij}\mathbb{R}^J_j$.  It is easy to verify that these are the desired CKVs: $\mathbb{D}_{<i}\mathbb{B}_{j>}^J=0=\mathbb{D}_{<i}\mathbb{R}_{j>}^J$, and that their commutators yield a representation of the Lorentz algebra.  For completeness we also state that $\mathbb{D}\cdot\mathbb{B}^J=-2r^J$ and $\mathbb{D}\cdot\mathbb{R}^J=0$, so we see from equation~(\ref{eq:Fequation}) that the boost generators contribute to $F$ (the radial perturbation) whereas the rotation generators do not, as expected.

Thus, the most general infinitesimal perturbation about the trivial RQF in flat spacetime is given by
\begin{eqnarray}
f^i(t,\theta,\phi) & = & \alpha^{I}(t)\,\mathbb{B}_{I}^{i}(\theta,\phi) + \beta^{I}(t)\,\mathbb{R}_{I}^{i}(\theta,\phi)\label{eq:fi}\\
F(t,\theta,\phi) & = & \alpha^{I}(t)\,r_{I}(\theta,\phi),
\end{eqnarray}
where $\alpha^{I}(t)$ and $\beta^{I}(t)$ are six arbitrary functions of time.  Substituting these expressions into equation~(\ref{eq:fIdecomposition}) yields
\begin{equation}
f^{I}(t,\theta,\phi)=\alpha^{I}(t)+\epsilon^{I}_{\phantom{I}JK}r^{J}(\theta,\phi)
\beta^{K}(t),\label{eq:fI}
\end{equation}
where $\epsilon_{IJK}$ is the alternating symbol, and we made use of the identity $\mathbb{S}^{ij}\,\mathbb{B}_{i}^{I}\,\mathbb{R}_{j}^{J}
=-\epsilon^{IJ}_{\phantom{IJ}K}\,r^{K}$.  By inspection of equation~(\ref{eq:GeneralPerturbation}) it is clear that $\alpha^{I}(t)$ and $\beta^{I}(t)$ correspond to time-dependent translations and rotations, respectively.

To understand the corresponding proper acceleration and twist the observers experience we begin by substituting equation~(\ref{eq:fi}) into equation~(\ref{eq:Perturbedu_i}):
\begin{equation}
u_i = \lambda r^2 \,[\, \mathbb{D}_i ( \,\dot{\alpha}_{I}r^{I}-\frac{c}{r}f^{0}   ) - \mathbb{E}_{i}^{\phantom{i}j}\mathbb{D}_j ( \dot{\beta}_{I}r^{I} ) ] + O(\lambda^2).
\end{equation}
We are free to choose $f^0$, so let us choose $\frac{c}{r}f^0=\dot{\alpha}_{I}r^{I}$, which is sufficient to make the exact (gradient) part of $u_i$ vanish, leaving only a co-exact (curl) part.  This is the part responsible for the failure of the time foliation to be hypersurface orthogonal.  We then have
\begin{eqnarray}
N & = & 1+\lambda \frac{r^2}{c^2} \ddot{\alpha}_{I}r^{I}  +O(\lambda^2)\\
u_i & = & \lambda r^2 \dot{\beta}_{I}\mathbb{R}^{I}_{i} + O(\lambda^2).
\end{eqnarray}
Comparing with $N=1+\frac{r}{c^2} r^{I} A_{I}$ from {\S}\ref{SimpleExamples-TranslationOnly} we see that the perturbed solution effectively has an acceleration parameter $A_I=\lambda r \ddot{\alpha}_I$, proportional to the second time derivative of the translation parameter.

In physical (coordinate covariant) terms, the observers' proper acceleration and twist are found using equations~(\ref{eq:ObserversAcceleration}) and (\ref{eq:nu}), and turn out to be
\begin{eqnarray}
\alpha_i &=& \lambda r^2\,( \ddot{\alpha}_{I}\mathbb{B}^{I}_{i} + \ddot{\beta}_{I}\mathbb{R}^{I}_{i} )+O(\lambda^2)\\
n\cdot a &=& \lambda r \,\ddot{\alpha}_{I}r^{I}+O(\lambda^2)\\
\nu &=& -\lambda \, \dot{\beta}_{I}r^{I}+O(\lambda^2).
\end{eqnarray}
(Again, the result for the normal acceleration, $n\cdot a$, is stated without proof.)  From the first equation we see that a nonzero second time derivative of the translation parameter $\alpha_I$ (respectively, rotation parameter $\beta_I$) is associated with a tangential proper acceleration having the pattern of a boost diffeomorphism generator, $\mathbb{B}_{I}^{i}$ (respectively, rotation diffeomorphism generator, $\mathbb{R}_{I}^{i}$).  The second and third equations tell us that translation (only) is associated with a proper acceleration normal to $\mathcal B$, and that rotation (only) is associated with a twisting congruence, at least in this lowest order (linear) approximation.

These results are intuitively sensible, and show that the most general RQF in the solution space neighborhood of the trivial RQF in flat spacetime has precisely the same six time-dependent degrees of freedom of motion as a rigid body in Newtonian space-time.  Moreover, these degrees of freedom can be identified with time-dependent Lorentz transformations acting on our quasilocal frame of observers via the CKVs, as mentioned above.  It is plausible---by continuity---that these properties are true for {\it general} RQFs (in both flat and curved spacetimes), however we have not proved this.  In the remainder of this section we will point out some subtleties involved in the general case.

Returning to the general notation introduced at the beginning of {\S}\ref{Definition} (and taking $c=1$) suppose that we have constructed an RQF in a generic spacetime, $(M,g_{ab})$.  This means we have a hypersurface ${\mathcal B}\approx \mathbb{R}\times \text{S}^2$ with spatial unit normal vector field $n^a$, and timelike unit tangent vector field $u^a$, such that the quotient space of the congruence (the space of integral curves of $u^a$) admits a well defined two-metric, $\sigma_{ab}$.  A perturbation of the RQF means a perturbation of the congruence, but this is equivalent to an infinitesimal active diffeomorphism of the spacetime metric, $g_{ab}$, in the neighborhood of $\mathcal B$, leaving the congruence fixed.  Let $\psi^a$ denote an arbitrary infinitesimal vector field defined in the neighborhood of $\mathcal B$ that effects this diffeomorphism: $\delta g_{ab} = 2\nabla_{(a}\psi_{b)}$, where $\nabla_a$ is the covariant derivative operator associated with the original spacetime metric.  On $\mathcal B$ we decompose $\psi^a$ as $\psi^a = \chi u^a + \Phi n^a +\phi^a$, where $\phi^a$ is tangent to the horizontal subspace, $H$, defined at the beginning of {\S}\ref{Definition}.  A simple exercise then shows that the corresponding change in $\sigma_{ab}$ is given by
\begin{equation}\label{eq:GeneralRQFPerturbation}
\delta\sigma_{ab}=2\left(\chi\theta_{(ab)}+\Phi k_{ab} + \tilde{\mathcal D}_{(a}\phi_{b)}\right),
\end{equation}
where $k_{ab}:=\sigma_{a}^{\phantom{a}c}\sigma_{b}^{\phantom{a}d} \nabla_{c}n_{d}$ is the $H$-projection of the (symmetric) extrinsic curvature of $\mathcal B$, and $\tilde{\mathcal D}_{a}$ is the covariant derivative operator induced on $H$, i.e., $\tilde{\mathcal D}_{a}\phi_{b}:= \sigma_{a}^{\phantom{a}c}\sigma_{b}^{\phantom{a}d} \nabla_{c}\phi_{d}$.  By assumption we have an RQF to begin with, so $\theta_{(ab)}=0$ and the first term on the right hand side of equation~(\ref{eq:GeneralRQFPerturbation}) vanishes, which tells us that the observers' two-metric is independent of the time reparametrization, $\chi$, as expected.

We now demand that the perturbation leave the observers' two-metric invariant, $\delta\sigma_{ab}=0$, i.e., while the observers may be in a different state of motion, they measure the same nearest-neighbor distances as in the original RQF. Taking the trace and trace-free parts of equation~(\ref{eq:GeneralRQFPerturbation}) (with respect to $\sigma_{ab}$) yields three linear partial differential equations
\begin{eqnarray}
\Phi & = & -\frac{1}{k}\,\tilde{\mathcal D}\!\cdot\! \phi\label{eq:Phi_equation}\\
\tilde{\mathcal D}_{<a}\phi_{b>}&=&\frac{k_{<ab>}}{k}\tilde{\mathcal D}\!\cdot\! \phi\label{eq:phi_equation},
\end{eqnarray}
in the three unknown functions $\Phi$ and $\phi^a$.  Here $k:=\sigma^{ab}k_{ab}$ and $\tilde{\mathcal D}\!\cdot\! \phi:= \sigma^{ab}\tilde{\mathcal D}_{a}\phi_{b}$.  As for the analogous equation~(\ref{eq:Fequation}), the normal perturbation, $\Phi$, is determined by the $H$-perturbation, $\phi^a$.  The problem is to determine the general solution to equation~(\ref{eq:phi_equation}) and show, in analogy with equation~(\ref{eq:CKVequation}), that the solution space is spanned by six arbitrary functions of time.

The principle subtlety here is that when the congruence is not hypersurface orthogonal the derivative operator $\tilde{\mathcal D}_a$ necessarily involves a time derivative. To see this explicitly, let us introduce, as before, a coordinate system $x^\mu :=(t,x^i)$ adapted to the congruence.  It can be shown that the $x^i$ coordinate components of equation~(\ref{eq:phi_equation}) are
\begin{eqnarray}
{\mathcal D}_{<i}\phi_{j>} & = & \frac{k_{<ij>}}{k}\,\left[  {\mathcal D}\!\cdot\! \phi+U^{k}\left( 2\nu\epsilon_{k}^{\phantom{k}l}\phi_{l}+\frac{1}{N}\dot{\phi}_k \right) \right]\nonumber\\
&  & -u_{<i}\left( 2\nu\epsilon_{j>}^{\phantom{j}k}\phi_{k} + \frac{1}{N}\dot{\phi}_{j>} \right) \label{eq:phi_equation_ij},
\end{eqnarray}
where the quantity in square brackets is $\tilde{\mathcal D}\!\cdot\! \phi$.  Here $U^{i}:=\sigma^{ij}u_{j}$ and ${\mathcal D}_i$ is the covariant derivative operator associated with the (time-independent) metric $\sigma_{ij}$, which is conformally related to the the covariant derivative operator ${\mathbb D}_i$ used earlier in this section (recall that $\sigma_{ij}=\Omega^2 {\mathbb S}_{ij}$).  In that case---perturbations about the trivial RQF in flat spacetime---$\nu$, $u_i$, and $k_{<ij>}$ are zero, and $k=2/r$.  Equations~(\ref{eq:Phi_equation}) and (\ref{eq:phi_equation_ij}) then reduce to (\ref{eq:Fequation}) and (\ref{eq:CKVequation}) and we recover our previous results, with $\Phi$ and $\phi_i$ being simply related to $F$ and $f_i$ by the constant conformal factor $\Omega=r$.

To gain some insight into the structure of equation~(\ref{eq:phi_equation_ij}), let us begin again with the trivial RQF in flat spacetime, except with a non-round bounding box, i.e., $\Omega \not=$ constant.  Then $u_i =0$, but $\hat{k}_{<ij>}:=k_{<ij>}/k\not=0$ and we need to solve the equation ${\mathcal D}_{<i}\phi_{j>} = \hat{k}_{<ij>}\,{\mathcal D}\!\cdot\! \phi$.  If we begin by setting $\phi_j$ equal to an arbitrary time-dependent linear combination of the six CKVs of the non-round bounding box (which are just the CKVs of the round sphere multiplied by $\Omega^2$), the left hand side is zero but the right hand side is, in general, not zero.  However, $\hat{k}_{<ij>}$, and the right hand side in general, is necessarily a linear combination of $\ell=2$ or higher (symmetric trace-free) tensor spherical harmonics.  So for the left hand side to equal the right hand side we must add to the ($\ell=1$) CKV part of $\phi_j$ vector spherical harmonics higher order in $\ell$.  But this in turn generates still higher order terms on the right hand side, and the process must be iterated, ultimately generating an infinite hierarchy of coupled linear equations for the time-dependent coefficients in a vector spherical harmonic expansion of $\phi_j$.  The key point is that this hierarchy of equations places no constraints on the starting ``seed"---the arbitrary time-dependent linear combination of six CKVs.  Thus, if for each such seed, the infinite hierarchy of equations yields a convergent, unique solution, we recover precisely six time-dependent degrees of freedom of motion for a non-round rigid box, the same as in Newtonian space-time.  It is certainly plausible that this is the case, however a proof is still in progress.

The more important case to consider is when the congruence is not hypersurface orthogonal, in which case $\nu$ and $u_i$ are necessarily non-vanishing.  Beginning with a CKV seed as before, equation~(\ref{eq:phi_equation_ij}) again generates an infinite hierarchy of coupled linear equations for the time-dependent coefficients in a vector spherical harmonic expansion of $\phi_j$.  The difference is that, because of the time derivative of $\phi_j$ occurring on the right hand side of the equation, the coefficients for higher $\ell$ will depend on higher order time derivatives of the coefficients in the CKV seed---in principle continuing up to infinite order.  This introduces a ``nonlocality" in time in the sense that the solution for $\phi_j$ (and thus also $\Phi$) on any time slice depends not only on the CKV data on that slice, but also an arbitrary number of time derivatives of this data.  Roughly speaking, this suggests that while we are free to specify how the bounding box is to accelerate and tumble as time passes (encoded in the CKV data), the higher order spherical harmonics that contribute to precisely locating the observers in space at any instant of time depend, in principle, on the entire history of the ($\ell=1$ component of the) box's specified motion.  In other words, unlike in Newtonian space-time, we cannot simply specify the position and velocity of the observers at an initial instant of time, and then integrate given their acceleration.  Again, a detailed analysis of the linearized RQF equations, (\ref{eq:Phi_equation}) and (\ref{eq:phi_equation}), is still in progress.

In the next section we will dispense with these linearized rigidity equations and solve, to the first few orders in a certain perturbation expansion, the full nonlinear equations in a highly nontrivial example involving a finite time-dependent rotation.  We will see the aforementioned nonlocality in time emerge.  This example also provides evidence that the hierarchy of equations in the linearized case should, indeed, yield a convergent, unique solution.

\section{\label{GeneralRotation}RQF Resolution of Ehrenfest's Paradox}

In this section we will provide a quasilocal resolution to the famous ``Ehrenfest's paradox," in which a rigid body at rest can never be brought into uniform rotation.\cite{EinsteinStudies-RotatingDisk}  In a certain perturbation expansion we will construct an RQF representing a round sphere in flat spacetime that begins at rest, is spun up to a relativistic angular velocity, and is then brought to rest again.  Roughly speaking, this means that concentric shells of a body can be spun up rigidly, but rigidity between neighboring shells cannot be maintained.

In {\S}\ref{SimpleExamples-ConstantRotationOnly} we considered a round sphere in flat spacetime rotating at a constant rate, $\omega$.  As a first approximation to a nonconstant rotation rate, $\omega(t)$, we take the ansatz in equation~(\ref{eq:ConstantRotationAnsatz}), with $\rho(\theta)$ and $z(\theta)$ replaced by $\rho(t,\theta)$ and $z(t,\theta)$ given by equations~(\ref{eq:rho}) and (\ref{eq:z}) with $\gamma$ replaced by $\gamma(t):=r\omega(t)/c$.  This is {\it not} an RQF, but if $\gamma(t)$ and its time derivatives are sufficiently small, it is a good approximation.  To this approximate solution we add a perturbation to the spatial embedding part (leaving the time foliation the same):
\begin{equation}\label{eq:TimeDependentRotationAnsatz}
\begin{array}{lll}
T & = & t\\
P & = & \rho(t,\theta)+\delta P(t,\theta)\\
\Phi & = & \phi + \int_0^t\omega(\tilde{t})\,\text{d}\tilde{t}+
\delta\Phi(t,\theta)\\
Z & = & z(t,\theta)+\delta Z(t,\theta),
\end{array}
\end{equation}
where $\delta P$, $\delta\Phi$, and $\delta Z$ are ``small" arbitrary functions, and we have replaced $\omega t$ with $\int_0^t\omega(\tilde{t})\,\text{d}\tilde{t}$.  We then compute the corresponding induced metric, $\gamma_{\mu\nu}$, in particular $\sigma_{ij}$, and demand that $\sigma_{ij}=r^2\mathbb{S}_{ij}$.  This results in three algebraic or differential equations for the three unknown functions, which we solve iteratively in powers of $\gamma(t)$ and its time derivatives.  (In reference to the previous section, the unperturbed congruence---which is not an RQF---is analogous to the CKV ``seed" specifying how the box is to rotate, and the perturbation is analogous to the higher order spherical harmonic corrections required to achieve this motion in a manner that maintains relativistic rigidity.)

For example, at the lowest order we find the three equations:
\begin{eqnarray}
0 &=:& \sigma_{\theta\theta}-r^2 \label{eq:sigma22}\\ &=& r^2\,\left\{ 2\cos\theta\frac{\delta P^\prime}{r} - 2\sin\theta\frac{\delta Z^\prime}{r} +\frac{\tau^2}{4}\sin^2{2\theta}  \gamma^2(t)\dot{\gamma}^2 (t)     \right\}\nonumber\\
0 &=:& \sigma_{\theta\phi}=r^2\sin^2\theta\left\{  \delta\Phi^\prime-\tau\sin\theta\cos\theta\, \gamma^2(t)\dot{\gamma}(t)  \right\}\label{eq:sigma23}\\
0 &=:& \sigma_{\phi\phi}-r^2\sin^2\theta \nonumber \\
&=& 2r^2\sin^2\theta \left\{  \frac{\delta P}{r} + \tau\sin^3\theta\,\gamma(t)\,\delta\dot{\Phi} \right\}
\label{eq:sigma33}
\end{eqnarray}
where, as before, a prime denotes differentiation with respect to $\theta$, and $\tau := r/c$ is the characteristic time for light to cross the system.  Solving equation~(\ref{eq:sigma23}) for $\delta\Phi$ with Dirichlet boundary conditions at $\theta=0$ and $\pi$ yields
\begin{equation}\label{eq:SolutionForDeltaPhi}
\delta\Phi(t,\theta)=\frac{1}{2}\tau\sin^2\!\theta\,\gamma^2(t)\dot{\gamma}(t).
\end{equation}
Substituting this result into equation~(\ref{eq:sigma33}) and solving for $\delta P$ yields
\begin{equation}\label{eq:SolutionForDeltaP}
\delta P(t,\theta) = -\frac{1}{2}r\tau^2\,\sin^5\!\theta\,\gamma^2(t) \left[
2\dot{\gamma}^2(t)+\gamma(t)\ddot{\gamma}(t)           \right].
\end{equation}
Finally, substituting this result into equation~(\ref{eq:sigma22}) and solving for $\delta Z$, with the condition that $\delta Z$ be an odd function about $\theta=\pi/2$, yields
\begin{eqnarray}\label{eq:SolutionForDeltaZ}
\delta Z(t,\theta) = \frac{1}{2}r\tau^2\,\cos^3\!\theta\,\gamma^2(t) \left[\,
\left(3-2\cos^2\!\theta\right)\dot{\gamma}^2(t)\right. \nonumber \\ \left.+\left( 5/3-\cos^2\!\theta \right)\gamma(t)\ddot{\gamma}(t)           \right].
\end{eqnarray}

From equation~(\ref{eq:SolutionForDeltaPhi}) we see that an angular acceleration, $\dot{\gamma}(t)\not= 0$, requires the set of observers on any given meridian of the rotating round sphere to suffer an azimuthal displacement that ``leads" the acceleration; this {\it bending} of the meridian lines in the azimuthal direction, $\delta\Phi\propto \sin^2\!\theta$, is seen by the {\it inertial} observers only, not the co-rotating observers.  For this to not distort the shape of the round sphere as seen by the co-rotating observers, the shape of the embedded sphere in the inertial frame must change.  This is accounted for by the terms proportional to $\dot{\gamma}^2(t)$ in $\delta P(t,\theta)$ and $\delta Z(t,\theta)$: relative to figure~1 there is an additional pulling in near the equator, and pushing out near the poles.

The most intriguing aspect of this perturbation away from the constant angular velocity solution is that in general it does {\it not} vanish when $\dot{\gamma}(t)= 0$: there are terms proportional to $\gamma^3(t)\ddot{\gamma}(t)$ in $\delta P(t,\theta)$ and $\delta Z(t,\theta)$.  For instance, consider beginning with a round sphere at rest, spinning it up to some angular velocity parameter $\gamma(0)>0$ at, say, $t=0$, and then bringing it to rest again, in a time-symmetric fashion so that $\dot{\gamma}(0)=0$ but $\ddot{\gamma}(0)<0$.  Although at $t=0$ the sphere is spinning with an angular velocity that is momentarily constant, the shape of the sphere as seen by the inertial observers is {\it not} that of a sphere {\it eternally} rotating with the same angular velocity parameter ($\gamma(0)$ in both cases): relative to figure~1 there is an additional pushing out near the equator, and pulling in near the poles.  (Note, however, that the bending of the meridian lines, $\delta\Phi$, {\it does} vanish at the time-symmetric point, as might be expected.)  The origin of this peculiar behaviour is the dependence of $\delta P$ on $\delta\dot{\Phi}$ (the time {\it derivative} of $\delta\Phi$---see equation~(\ref{eq:sigma33})), which can be traced back to the nonlinear $u_i u_j$ term in the relation $\sigma_{ij}=\gamma_{ij}+\frac{1}{c^2}u_i u_j$, and is also related to the $u_{<i}\dot{\phi}_{j>}$ term on the right hand side of equation~(\ref{eq:phi_equation_ij}).

Moreover, using GRTensor II running under Maple it is not difficult to iterate this perturbation expansion to higher powers in $\gamma(t)$ and its derivatives.  Indeed, we have iterated ``two and a half" more times, up to terms in $\delta\Phi(t,\theta)$ involving the seventh time derivative of $\gamma(t)$, and there appears to be no obstruction to continuing the iterations indefinitely, except that the expressions grow in size exponentially with successive iterations.  Thus it seems that the shape of the sphere at any instant of time, $t$, as seen by the inertial observers, depends not only on $\gamma(t)$ at that instant, but also on all of its time derivatives up to infinite order, i.e., it depends on the entire history of $\gamma(t)$.  This is the ``nonlocality" in time discussed in the context of the linearized rigidity equations in the previous section.  We did not see this behaviour in the case of finite time-dependent translation (with no rotation) because in that case the congruence is hypersurface orthogonal, so at worst $u_i$ is exact (a gradient); in fact, we used a time foliation of $\mathcal B$ for which $u_i$ simply vanishes.  For the case of finite time dependent rotation the twist is nontrivial and, according to equation~(\ref{eq:nu}), $u_i$ contains an irremovable co-exact (curl) part.  And then the $u_i u_j$ term in the relation $\sigma_{ij}=\gamma_{ij}+\frac{1}{c^2}u_i u_j$---which is present because of the relativity of simultaneity---comes into play.  In the end, though, this nonlocality in time is perhaps not surprising if one recalls that rigid motion already involves nonlocality in space in the sense that the acceleration at one point on the bounding box depends on the accelerations at causally disconnected points: equation~(\ref{eq:CKVequation}) is an elliptic differential equation.

As an example to test the robustness of our solution, we considered the particular choice of time dependent angular velocity given by
\begin{equation}
\omega \left( t \right) =\Omega\,{e^{-{{t}^{2}/{T}^{2}}}},
\end{equation}
that describes a time-symmetric situation in which the sphere spins up to a maximum angular velocity $\Omega$ and then back down to zero with a time scale $T$.  Considering both the absolute and relative magnitudes of the successively higher order terms in the solution, we can, with confidence, push our approximate solution to the rather extreme case of $\Omega\approx 1/3$ and $T\approx 1$ (with $r=1=c$).  This corresponds to observers on the sphere's equator spinning up to $v_\text{equator}\approx c/3$ (see equation~(\ref{eq:vEquator})) on a time scale about equal to the time it takes light to cross the system, in this case a small fraction of one revolution.

\section{\label{Conclusions}Conclusions}

Rigid motion in Newtonian space-time has six time-dependent degrees of freedom: three translations and three rotations.  Rigid motions also exist in both special and general relativity, but they are severely restricted, as outlined in the first paragraph of the Introduction section.

In this paper we have introduced the concept of a {\it rigid quasilocal frame} (RQF), which opens up the possibility of rigid motion in both special and general relativity with the full six time-dependent degrees of freedom we have in Newtonian space-time.  The definition of an RQF, applicable in both flat and curved spacetimes, is identical to Born's definition of rigidity,\cite{Born1909} except the key is to consider not a three-parameter congruence of timelike worldlines (a swarm of observers filling a three-dimensional volume of space), but a two-parameter congruence (the observers on the topologically $S^2$ boundary of the volume---a {\it quasilocal} frame).

As proofs of principle we have constructed, either exactly or approximately, several examples of RQFs in flat spacetime, including one that directly addresses the difficult problem of finite time-dependent rotations---see previous section: RQF Resolution of Ehrenfest's Paradox.  Two key aspects of quasilocal rigidity emerged.  The first is that the existence of six degrees of freedom in the rigid motion of our two-sphere's worth of observers is intimately connected with the fact that any two-surface with the topology $S^2$ always admits precisely six conformal Killing vectors (CKVs), which generate a representation of the Lorentz algebra.  In contrast to the usual case of the Lorentz group acting locally on a single observer (rotations and boosts of his tetrad along his worldline), here we have the Lorentz group acting {\it quasi}locally on a two-sphere's worth of observers along their worldtube.

The second is that finite time-dependent rotations are subtle precisely because they introduce a ``nonlocality" in time: unlike in Newtonian space-time it is not possible to specify a cross section of the (twisting) congruence in space, on a given time slice, without knowing the entire history of the motion.  Of course RQFs are also nonlocal in space in the sense that rigid motion requires observers at different points on the boundary to act in concert.  Thus an RQF is an inherently nonlocal construction in spacetime.  (It is perhaps worth emphasizing that in any given spacetime, an RQF is simply a congruence of worldlines with certain geometrical properties.  Throughout this paper the word ``observer" is used in an abstract sense.  We are not requiring or implying that a two-sphere's worth of physical observers in a dynamical spacetime can actually adjust their accelerations with physical thrusters in concert to achieve an RQF.  Also, the existence of RQFs does not require or imply the existence of rigid bodies.)

A large set of results on RQFs exists, generated by various of the present authors, which lie  outside the scope of this paper and which will appear in future publications.  For example, the case of a round sphere RQF in a generic spacetime with matter has been analyzed as a series expansion in powers of the areal radius, $r$, similar to the construction of Fermi normal coordinates.  The analysis---carried out to the first few orders in $r$---indicates that the transition from flat to curved spacetime introduces no obvious obstructions; we still obtain six time-dependent degrees of freedom.  Moreover, as alluded to in the abstract, the application of RQFs to the problem of motion and the nature of gravitational radiation have yielded several interesting new insights.

In addition to exploring more applications, it is important that the existence and uniqueness (up to six degrees of freedom) of solutions to the RQF rigidity conditions be rigorously proven.  Exact solutions of at least some representative examples of nontrivial RQFs would also be extremely useful.  Work in these directions is in progress.

\bigskip
\section*{Acknowledgements}
R.J.E. would like to thank Luis de Menezes for discussions.  This work was supported in part by the Natural Sciences and Engineering Research Council of Canada.   The authors would like to thank the referees for comments that led to improving the clarity of the paper at several points.

\end{document}